# Small variation in dynamic functional connectivity in cerebellar networks

Izaro Fernandez-Iriondo[a], Antonio Jimenez-Marin[a,b], Ibai Diez[c,d], Paolo Bonifazi[a,e], Stephan P. Swinnen[f,g], Miguel A. Muñoz[h,j] and Jesus M. Cortes[a,e,i,j]

[a]*Biocruces-Bizkaia Health Research Institute, Barakaldo, Spain*
[b]*Biomedical Research Doctorate Program, University of the Basque Country (UPV/EHU), Leioa, Spain*
[c]*Functional Neurology Research Group, Department of Neurology, Massachusetts General Hospital, Harvard Medical School, Boston, MA 02115, USA*
[d]*Gordon Center, Department of Nuclear Medicine, Massachusetts General Hospital, Harvard Medical School, Boston, MA 02115, USA*
[e]*IKERBASQUE: The Basque Foundation for Science, Bilbao, Spain*
[f]*Movement Control and Neuroplasticity Research Group. KU Leuven, Leuven, Belgium*
[g]*Leuven Brain Institute (LBI). KU Leuven, Leuven, Belgium*
[h]*Instituto de Física Teórica y Computacional Carlos I, Universidad de Granada, Facultad de Ciencias, Granada, Spain*
[i]*Departament of Cell Biology and Histology. University of the Basque Country (UPV/EHU), Leioa, Spain*
[j]*Equal last-author contribution*

Abstract

Brain networks can be defined and explored through their connectivity. Here, we analyzed the relationship between structural connectivity (SC) across 2,514 regions that cover the entire brain and brainstem, and their dynamic functional connectivity (DFC). To do so, we focused on a combination of two metrics: the first assesses the degree of SC-DFC similarity –i.e. the extent to which the dynamic functional correlations can be explained by structural pathways–; and the second is the intrinsic variability of the DFC networks over time. Overall, we found that cerebellar networks have a smaller DFC variability than other networks in the brain. Moreover, the internal structure of the cerebellum could be clearly divided in two distinct posterior and anterior parts, the latter also connected to the brainstem. The mechanism to maintain small variability of the DFC in the posterior part of the cerebellum is consistent with another of our findings, namely, that this structure exhibits the highest SC-DFC similarity relative to the other networks studied, i.e. structure constrains the variation in dynamics. By contrast, the anterior part of the cerebellum also exhibits small DFC variability but it has the lowest SC-DFC similarity, suggesting a different mechanism is at play. Because this structure connects to the brainstem, which regulates sleep cycles, cardiac and respiratory functioning, we suggest that such critical functionality drives the low variability in the DFC. Overall, the low variability detected in DFC expands our current knowledge of cerebellar networks, which are extremely rich and complex, participating in a wide range of cognitive functions, from movement control and coordination to executive function or emotional regulation. Moreover, the association between such low variability and structure suggests that differentiated computational principles can be applied in the cerebellum as opposed to other structures, such as the cerebral cortex.

## Introduction

Understanding the relationship between different classes of connectivity is fundamental in network neuroscience [1]. To date, different strategies exist to obtain structural connectivity (SC) matrices from magnetic resonance imaging (MRI), where each entry to the matrix represents the number of white-matter streamlines between pairs of brain regions obtained from diffusion-weighted images (DWIs) [2]. Similarly, there is considerable information as to how to construct functional connectivity (FC) matrices, obtained by assessing the similarity in the dynamics of given pairs of brain regions from a sequence of functional images employing diverse metrics, e.g., pairwise Pearson correlations or synchronization measures [2]. However, despite these significant advances, a protocol to define or predict one connectivity class from another remains to be defined.

One issue that complicates matching SC to FC is the fact that they involve very different time scales even though they deal with network connectivity between the same brain regions. Hence, SC is practically invariant in the period over which the FC is calculated (typically a maximum of 10 minutes), the latter known to vary over short time scales of even a few seconds, exhibiting a rich dynamic repertoire (see [3] and references therein). When considering short time scales, the simplest manner to assess and quantify the temporal variation in "dynamic functional connectivity" (DFC) is by considering a sliding window analysis, where the total window length is divided in several intervals of a fixed duration. One FC is then calculated for each time window and in this way, the DFC is represented as a time ordered sequence of FC matrices.

Importantly, it has been shown that the SC can be inferred from the DFC when the time window to calculate it is infinitely large [4]. In other words, pairwise correlations –when averaged over sufficiently long time periods– merely reflect the underlying SC matrix (see also [5]). This fact has become even clearer through the observation that functional networks in the resting state can be derived from the spectrum of eigenmodes –or harmonics– of the SC matrix (actually, from its associated Laplacian matrix [6]). By contrast, when functional networks are obtained over much shorter time windows, the dynamics of the brain operating on a fixed SC network generates a large repertoire of different varying functional networks. However, how such dynamic patterns are related to the brain's function and disease remains largely unclear, despite the substantial advances achieved over recent years [4, 7–11].

---

*Corresponding author
✉ jesus.m.cortes@gmail.com (J.M. Cortes)
ORCID(s): 0000-0002-9059-8194 (J.M. Cortes)

The relationship between the temporally-invariant SC and the highly temporal-sensitive DFC can be assessed by comparing the two graphs at the level of individual links, a strategy which requires –for symmetrical matrices– $\mathcal{N}^2/2$ comparisons (where $\mathcal{N}$ is the number of nodes in the network). Alternatively, here we follow another and more efficient strategy that involves establishing a comparison at a modular or community level [12]. In particular, using a standard algorithm for community detection [13, 14], modules can be identified from either structural and/or functional matrices, and the two types of networks are then compared using the same module-representation for the two classes of networks. Our hypothesis is that if we assume that segregated functions are associated with distinct structural modules, visualizing the functional modules in terms of the structural ones should help define and highlight how strongly structural constraints affect function, and vice versa.

In this study, we assessed how SC constraints affect DFC at the module level, and we found that DFC cerebellar modules were much less variable than those in the cerebrum. We show that the small variability found is driven by two different mechanisms, one mediated by the constraint imposed by the SC and the other, possibly related to external influences that affect function. The small variability in the cerebellar DFC networks detected in this study might reflect that different computational principles are active in the cerebellum as opposed to other brain circuits.

## Materials and methods
### Participants

A population of healthy subjects ($\mathcal{P} = 48$) were recruited from the general population in the vicinity of Leuven and Hasselt (Belgium) through advertisements on websites, announcements at meetings and the use of flyers at visits of organizations and public gatherings (PI: Stephan Swinnen). The participant's ages ranged between 20 and 51 years (mean 33.9 and standard deviation 9.79), and none of the participants had a history of ophthalmological, neurological, psychiatric or cardiovascular diseases that could potentially influence the imaging or clinical measures. All the participants provided their informed consent before enrollment on the study, in agreement with the local ethics committee for biomedical research.

### Image acquisition

*Anatomical data:* A high-resolution T1 image was acquired with a 3D magnetization prepared rapid acquisition gradient echo (MPRAGE): repetition time (TR)= 2, 300 ms, echo time (TE)= 2.98 ms, voxel size = $1 \times 1 \times 1.1$ mm$^3$, slice thickness = 1.1 mm, field of view (FOV)= $256 \times 240$ mm$^2$, 160 contiguous sagittal slices covering the entire brain and brainstem.

*Diffusion weighted imaging (DWI):* A DWI SE-EPI (diffusion weighted single shot spin-echo echo-planar imaging [EPI]) sequence was acquired with the following parameters: TR = 8, 000 ms, TE= 91 ms, voxel size = $2.2 \times 2.2 \times 2.2$ mm$^3$, slice thickness = 2.2 mm, FOV = $212 \times 212$ mm$^2$, 60 contiguous sagittal slices covering the entire brain and brainstem. A diffusion gradient was applied along 64 non-collinear directions with a $b$ value of 1, 000 s/mm$^2$. Additionally, one set of images was acquired without diffusion weighting ($b = 0$ s/mm$^2$).

*Resting state functional data:* Acquired with a gradient EPI sequence over a 10 min session using the following parameters: 200 whole-brain volumes with TR/TE = 3, 000/30 ms, flip angle = 90°, inter-slice gap = 0.28 mm, voxel size = $2.5 \times 3 \times 2.5$ mm$^3$, $80 \times 80$ matrix, slice thickness = 2.8 mm, 50 oblique axial slices, interleaved in descending order. During resting state acquisition, all participants received the instructions to keep their eyes open and to not think of anything in particular.

### Image preprocessing

*Diffusion images:* We applied a pre-processing pipeline similar to that employed previously [15–22] using the FSL (FMRIB software Library v5.0) and the Diffusion Toolkit. First, an eddy current correction was applied to overcome the artifacts produced by variation in the direction of the gradient fields in the MR scanner, together with the artifacts produced by head movements. Specifically, the participant's head movement was extracted from the transformation applied at the step of eddy current correction. The movement information was also used to correct the gradient directions prior to tensor estimation. From the corrected data, a local fitting of the diffusion tensor per voxel was obtained using the *dtifit* tool incorporated into FSL and finally, fiber assignment was achieved with a continuous tracking algorithm [23]. We then computed the transformation from the Montreal Neurological Institute (MNI) space to the individual-participant diffusion space and chose the network nodes to calculate the SC using a functional partition (see below).

*Functional images:* We applied a pre-processing pipeline similar to previous work [15–18, 24–26] using FSL AFNI (http://afni.nimh.nih.gov/afni/). A slice-time correction was first applied to the fMRI data set and then, the each volume was aligned to the middle volume to correct for head movement artifacts. After intensity normalization, we regressed out the movement time courses, the average cerebrospinal fluid (CSF) signal, the average white matter signal and the global signal[1]. A bandpass filter was then applied between 0.01 and 0.08 Hz [27], and the preprocessed functional data were spatially normalized to the MNI152 brain template, with an isotropic[2] voxel size of 3 mm. All the voxels were then spatially smoothed with a 6 mm full width at half maximum isotropic Gaussian kernel. Finally, and in addition to head movement correction, we performed scrubbing, through which time points with a frame-wise displacement > 0.5 were interpolated with a cubic spline [28]. We finally removed the effect of head movement using the global frame displacement as a covariate.

### Calculation of SC and DFC

Both the SC and DFC were built using $\mathcal{N} = 2,514$ nodes, identified after running an unsupervised method to cluster the functional data [29]. On average, each cluster –which in this study coincides with one node in the network– contained about 20 voxels of $3 \times 3 \times 3$ mm$^3$. One SC matrix of $2,514 \times 2,514$ dimensions was obtained for each subject by counting the number of white matter streamlines connecting all possible $2,514 \times 2,514$ pairs of nodes. Thus, the element matrix $(i, j)$ of SC is given by the streamline number between nodes $i$ and $j$, with $i, j = 1, \ldots, \mathcal{N}$. Given the lack of directionality of the streamlines the SC is a symmetric matrix. To calculate the population SC matrix, denoted by pSC, we first binarized the individual SC matrices and then took the overall average of the participants.

With respect to the functional networks, after averaging all the voxel time series within each network node, we extracted a single time series for each of the 2,514 nodes. By dividing the total time series length $\mathcal{T}$ in $\mathcal{W}$ non-overlapping windows of length $\delta$, we obtained $\mathcal{P} \times \mathcal{W}$ matrices, where $\mathcal{P}$ is the number of participants and $\mathcal{W}$ the number of time windows. DFC $\equiv \{FC_w\}_{w=1}^{\mathcal{W}}$ was defined as the temporal sequence of squared matrices $FC_w$, each one of dimension $\mathcal{N} \times \mathcal{N}$ and calculated over a fixed window $w$ by assessing the pairwise Pearson correlation coefficient between all-node time series within the time window $w$. The population DFCs, denoted by pDFC, were built by averaging each one for all the participants. Note that the DFC is a tensor, although we can also refer to the two objects DFC and $FC_w$ at the module level, simply extracting from them the within the module contributions, which we will denote as DFC$^m$ and $FC_w^m$, respectively. The former is another tensor composed of a sequence of matrices of dimensions $\mathcal{N}_m \times \mathcal{N}_m$ and the latter is a squared matrix of dimension $\mathcal{N}_m \times \mathcal{N}_m$. For both cases, $\sum_{m=1}^{M} N_m = N$.

---

[1] We also repeated our analyses without global signal regression.
[2] Notice that although the original functional voxel was not isotropic, however, the processed images are transformed into the MNI152 template, where voxels are now isotropic with a size of 3 mm.

### Adapting high-pass filtering for very short time windows

Conventional FC approaches typically work with long-time series. However, for relatively short time windows, the calculation of the $FC_w$ matrix requires adapting the lowest bound (LB) used for band-pass filtering of the time series using the equation $LB^{-1} = \delta \times TR$, where $\delta$ is the window length [30]. For example, for $\delta = 8$ and TR=3 seconds (as used here), the high-pass filter has to be adapted by taking LB = 0.042 rather than 0.01, as used here in the pipeline to pre-process the functional data.

### Structural modules used as a template for reordering DFC

Maximizing the modularity of the pSC matrix by employing the algorithm described in [31], we obtained a subdivision of the structural network into $\mathcal{M}$ non-overlapping modules that maximizes the number of within-module connectivity whilst minimizing that between modules [14]. We next used this partition to reorder the elements of the functional matrices, and to assess the link-to-link —or pairwise— similarity between the SC and $FC_w$ matrices. This comparison was repeated for all the functional matrices obtained in the different windows. In this way, we achieved a common structural organization of the modules, which was used as a template to reorder all the functional matrices. As found previously [12], this is a very convenient strategy to highlight the similarities and differences between both types of networks at a "mesoscopic" level.

### Assessment of SC-DFC similarity

After reordering all the functional networks using the structural modules, the SC-DFC similarity was assessed at the level of the modules. As such, we first extracted the $\mathcal{M}$ squared matrices pSC$^m$ and pFC$_w^m$ from the original pSC and pFC$_w$ matrices. For each module and window $w$ we calculated the Pearson correlation as a similarity measure: $r_w^m = \rho(\overrightarrow{pSC^m}, \overrightarrow{pFC_w^m})$, where $\overrightarrow{pSC^m}$ and $\overrightarrow{pFC_w^m}$ represent the vector-wise representation of matrices pSC$^m$ and pFC$_w^m$, respectively. Finally, we averaged the similarity over windows of the same size, i.e.: $r^m = <r_w^m>_w$.

### Assessment of DFC variability

For a given window length, we obtained a series of consecutive pFC$_w^m$ matrices using the $\mathcal{M}$ structural modules. For each of the $m = 1, \ldots, \mathcal{M}$ modules and $w = 1, \ldots, \mathcal{W}$ windows, we assessed the variability over the different time windows by calculating their pairwise spectral distance [32]:

$$\Delta_{w,w'}^m = \sum_{u=1}^{\mathcal{N}_m} \left| \lambda_u(pFC_w^m) - \lambda_u(pFC_{w'}^m) \right|, \qquad (1)$$

where $w, w' = 1, \ldots, \mathcal{W}$ are two generic windows, and $\lambda_1(G) \leq \lambda_2(G) \leq \lambda_3(G) \leq \ldots \leq \lambda_{N_m}(G)$ of the two graphs $G = \{pFC_w^m, pFC_{w'}^m\}$ are the sets of eigenvalues.

Note that other metrics can be used to assess the DFC variations, such as the temporal variance of the connectivity matrix across the time windows [33, 34] or its temporal standard deviation [35, 36]. Here, our choice relies on the spectral properties of the graphs, namely that the eigenvalues of the two isomorphic graphs are identical. Thus, Eq. (1) is one possible way to quantify the isomorphic-separability in pairs of graphs.

**Labelling of the anatomical regions**

The anatomical identification of each module (cf. Tables 1 and S1) was achieved by calculating the percentage overlap between each module and each region in the bi-lateralized automated anatomical labeling (AAL) brain atlas [37], pooling the left and right sides of each anatomical structure in the original atlas into the same region. Only regions with an overlap above 5% were reported.

**Cerebellar projections towards resting state networks**

To further understand the functional roles of the cerebellar modules, we projected them onto a highly specialized cerebellar atlas [38, 39], containing information about the projections from each cerebellar region in the atlas towards each resting-state network. We finally calculated the percentage overlap between our modules and each region in the cerebellar atlas.

**SC matrices using probabilistic tractography**

We also calculated the SC matrices using probabilistic tractography, as implemented in FSL. We first estimated a probabilistic model to compute the fiber orientation using bedpost [40]. We then calculated the connectivity matrices (one per subject) using the *probtrackx2* function and 100 pathways per voxel. Connectivity matrices were calculated by defining a threshold for the probability to define a non-zero connection such that both matrices, the population-deterministic one used for all the other analyses and the population-probabilistic one, had approximately the same number of non-zero elements, i.e.: the two matrices achieved the same link-density as when 97.6% of all probabilistic connections were fixed to zero.

**Overlapping time-windows for the calculation of DFC**

Although the results presented here considered non-overlapping time windows for the calculation of DFC, we also assessed DFC using sliding windows with an overlap of 96% between them [41].

**Templates for Resting State Networks**

We also calculated the two metrics analyzed here (DFC variability and SC-DFC similarity) for classic resting state networks (RSNs). In particular, we made use of the templates developed previously [42] for the four following RSNs: Default Mode Network (DMN), Ventral Attention Network (VAN), Dorsal Attention Network (DAN) and Somato-Motor Network (SMN).

**Synchronization metrics for DFC**

In addition, to assess DFC by calculating the pairwise Pearson's correlation between node time series, we also established synchronization metrics. These were calculated by first obtaining the Hilbert transformation of the node time series $x(t)$ represented by $\hat{x}(t) \equiv \mathcal{H}\{x\}(t)$ to build the complex analytical signal as $x(t) + i\hat{x}(t)$, represented as $A(t)\exp(i\theta(t))$, where $A(t)$ and $\theta(t)$ represent the instantaneous amplitude and the instantaneous phase of the analytical signal, respectively. We then obtained the $FC_w$ matrices in two more different forms: (1) calculating the pairwise Pearson's correlation between the time series of the instantaneous amplitude along different time windows; and (2) by calculating the phase locking value (PLV) [43–46], defined for any pair of nodes i and j as:

$$\text{PLV}^w(i,j) = \frac{1}{\delta}\left|\sum_{t \in w} e^{i(\theta_i(t) - \theta_j(t))}\right|, \quad (2)$$

where the average is taken in a time window of length r, and $|\cdot|$ indicates the modulus of a complex number. The PLV takes values in the interval [0,1], with 0 corresponding to the case where there is no phase synchrony and 1 whenever the two phases of the two signals are always identical.

## Results

A population of young healthy participants ($\mathcal{P} = 48$) was studied here, acquiring diffusion and resting-state images for each participant (see pipeline in figure 1). We first divided the population SC matrix, represented by the pSC, through modularity maximization, resulting in $\mathcal{M} = 14$ non-overlapping modules (see figure 2) with a modularity index of 0.7324. Modules 6 and 13 had 2 and 1 regions, respectively, and therefore, neither of these modules were considered in the following quantitative analyses.

After calculating one $pFC_w$ matrix per time window $w$, we reordered all of them according to the structural modules and calculated the link-to-link correlation for all the modules separately, $r_w^m$, as a measure of the similarity between the $pSC^m$ and the different $pFC_w^m$ (figure 2). In particular, we first considered $\delta = 8$, equivalent to a time duration of 24 seconds, and averaged the measurement across all the time windows.

Moreover, we studied the variability in $pFC_w$ across the time windows, in this case measuring the pairwise spectral distance $\Delta_{w,w'}^m$ within each module $m$ for the time windows $w$ and $w'$, and averaging this over all pairs $(w, w')$. The analysis of the SC-DFC similarity across the different modules revealed the existence of an outlier, module 1, which had a similarity value of $r^m > 0.7$, contrasting with the rest of the modules whose similarity was less than 0.55 (figure 3). The anatomy of module 1 is shown in Table 1 and strikingly, it is mainly formed by posterior cerebellar structures. By contrast, the lowest SC-DFC similarity was reported for module 14, with $r^m < 0.4$ for all time-windows. Importantly, this module is formed by a different part of the cerebellum (its anterior part: see Table 1), together with other non-cerebellar structures like the brainstem, the fusiform nucleus and part of the lingual cortex.

In addition to the SC-DFC similarity, we quantified the amount of DFC variability and found that module 11 exhibited the highest variability over time (Table 1), a module composed of several cortical structures that include the calcarine, middle and inferior temporal, lingual and precuneus.

By assessing both metrics simultaneously, $r^m$ and $\Delta^m$, modules 1 and 14 had considerably smaller $\Delta^m$ values than module 11 (figure 3A), indicating that the cerebellar structures displayed much less DFC variability over time than other structures in the cerebrum (figure 3A shows the mean values across windows and the histograms of all the possible values are shown in figure 3B). Several brain slices from these three modules are shown in Figure 3C and the anatomical composition of the rest of the modules analyzed is given in Table S1.

To further demonstrate the robustness of our findings, that cerebellar modules 1 and 14 fulfil a differentiated role in terms of DFC variability and SC-DFC similarity, we performed several control-analyses under different conditions that included varying some image-preprocessing steps, or using different parameters in our modelization or different metrics to calculate the DFC (see Methods for details). First, we repeated the same analysis but considering probabilistic rather than deterministic tractography (figure S1A). Second, we performed a sliding window analysis to calculate the DFC using 96% overlapping windows rather than non-overlapping windows (figure S1B). Third, we preprocessed all the functional images without global signal regression and repeated the same analyses (figure S1C). Finally, we calculated different metrics to assessing DFC variability but based on the analytical signal, which is therefore more closely related to standard synchronization studies. In particular, we assessed the DFC by calculating the pairwise correlations between time series of instantaneous amplitude (figure S1D) and instantaneous phase (figure S1E), that is, the PLV. Indeed, a similar equivalence between different ways to construct functional matrices has been also reported elsewhere [47]. In all these situations, both cerebellar modules 1 and 14 preserved their differential role relative to the other modules in the cerebrum.

We also asked whether our findings obtained for a window of $\delta = 8$ were robust when varying the length of the window (figure 4). Specifically, we obtained all the measurements again for the following window lengths: $\delta = \{4, 5, 7, 8, 10, 25\}$, and also for $\delta = \mathcal{T}$, equivalent to a single time window with a length equivalent to the entire time series. In this latter case, we only addressed the SC-DFC similarity because DFC variability could not be assessed in a single time window (see Table S2). Overall, we found two cerebellar modules that preserved their roles irrespective of the window length chosen and of other control conditions (figure S1), modules 1 and 14 in the posterior and anterior part of the cerebellum, respectively. By contrast, when we looked at module 11 over different window lengths, the module with the highest DFC variability, we noticed that this behavior was more variable across time windows and that modules 2 and 11 interchanged their positions.

We next asked which brain areas in modules 1 and 14 were structurally and functionally connected (Table 2), and we found strong mutual-connectivity between the anterior and posterior parts of the cerebellum, as seen previously [48, 49]. Moreover, the anterior region also projects to the motor cortex, as has been well established [50–52]. By quantifying the overlap of modules 1 and 14 with the cerebellar regions projecting to different RSNs (see Methods), we found that module 1 projected towards the DMN (22.8%) and the frontoparietal network (19.2%), while module 14 did so towards the SMN (17.7%). Moreover, both modules projected similarly to the VAN, with a 13.3% overlap for module 1 and 10.5% for module 14. Thus, these results show in a different manner that both cerebellar modules are different yet complementary to each other, with module 1 participating in high order cognitive networks and module 14 in somatomotor networks, while both participate in multimodal integration networks like those associated with ventral attention [53].

Finally, we assessed the SC-DFC similarity and DFC variability in classical RSNs [42, 54–56]. In particular, we analyzed the DMN, SMN, VAN and DAN networks (figure 5), which all had much less SC-DFC similarity than our modules. This probably reflects the fact that the RSNs were built purely from functional data, whereas we combined functional and structural data together to obtain the final modules. Moreover, we also found that the DMN had the highest DFC variability when compared with the other RSNs and also with our modules.

## Discussion

We have assessed here the relationship between SC and DFC by combining structural and functional brain networks, and assessing their modular organization. We have built networks with high spatial resolution, covering the entire brain, using 2, 514 nodes that each have an average size of 0.54 cm$^3$. Through modularity maximization of the population SC matrix, we obtained 14 non-overlapping structural modules that were used to reorder the functional matrices. This was done under the assumption that if segregated functions are associated with distinct structural modules, visualizing the functional matrices following their structural fingerprints would clarify how SC constrains function, a fundamental question that has yet to be resolved.

We analyzed the SC-DFC similarity in combination with the amount of DFC variability over time for all the previously identified modules, and characterized each module in terms of these two metrics. This allowed us to identify three ex-treme cases: 1, a fully cortical module with the highest DFC variability; 2, a module in the posterior cerebellum that had the highest SC-DFC similarity while maintaining low DFC variability; 3, a module in the anterior cerebellum connected to the brainstem that had the lowest SC-DFC similarity but also, that retained small DFC variability. Therefore, cerebellar networks appear to have small DFC variability. Module 1, located at the posterior cerebellum, has about twice the similarity between SC and DFC than the rest of the modules, indicating that structure constrains the dynamic connectivity [57, 58].

However, how module 14 associates weak SC-DFC similarity with low DFC variability is more challenging to understand. On the one hand, module 14 includes the brainstem in addition to the anterior part of the cerebellum, the former having strong connectivity to many other parts of the brain and body through major tracts like the corticospinal, lemniscus and spinothalamic tracts [59]. Thus, by looking at the intra-module similarities between SC and DFC, as performed here, it is possible that we ignored relevant aspects of connectivity from this module to the rest of the brain, thereby underestimating the intra-module SC-DFC similarity. Moreover, it is well known that the brainstem plays a critical role in regulating sleep cycles, as well as cardiac and respiratory function. Perhaps, such critical functions are not compatible with large DFC variability, as occurs in cortical networks, although future research will be needed to fully explain these observations. Future studies should also verify our results on larger sample sizes.

It is important to emphasize that in terms of SC-DFC similarity, our unsupervised method identifies two divisions within the cerebellum, the posterior and anterior regions. This turns out to be a well-known and standard anatomical and functional division of the cerebellum in humans and animals [48]. Moreover, while the classical cerebellum division grouped the anterior lobules from 1 to 5 [52], our results include lobule 6 in both the anterior and posterior cerebellum, in agreement with functional MRI studies that indicated cerebellar lobules 4-6 participate in sensorimotor tasks [60].

We found that module 1, the posterior cerebellum, participated in high-order cognitive functions, as reflected by a strong overlap with the DMN and the frontoparietal network, in agreement with previous data [61]. This is also consistent with the fact that the cerebellum areas Crus 1 and 2, included in module 1, connect to the thalamus and then, to prefrontal areas. In relation to module 14, the anterior part of the cerebellum, we found projections towards the somatomotor network, in agreement with [61]. Moreover, both these modules projected to multimodal integration hubs like the VAN [53].

In the resting state, the DFC is dominated by the DMN, together with attentional and sensory networks [62, 63]. We show here that DMN variability was indeed the highest of all the modules and networks studied, which might suggest why different DFC patterns in the DMN can encode multiple brain states [64], and also why the DFC is more limited in several pathological conditions [65]. Although our results were obtained in the resting state and therefore, the participants were not performing any specific task in the MRI scanner, these findings might also explain that when a motor task is more difficult to perform or it is simply a new task for the subject, posterior cerebellar activation occurs together with prefrontal activity to enhance cognitive monitoring of the individual's performance [60, 66–69].

It is well-known that cerebellar networks have an extremely rich and complex anatomy and functionality [70], connecting to the brainstem and cerebral hemispheres, and participating in a large variety of cognitive functions like movement coordination, bimanual coordination performance [71], executive function, visual-spatial cognition, language processing and emotional regulation [72, 73]. However, as far as we know the small variability of the DFC within the cerebellum has not been reported previously.

Finally, the extraordinary constraint of the DFC to the SC in cerebellar module 1 might indicate distinct operational and computational principles in the cerebellum. Classically, cerebellar architecture has been modeled in a feedforward manner, unlike the highly recurrent circuits found in the cerebral cortices (see [74] and references therein). This phenomenon enables the cerebellum to linearly integrate different inputs from other systems in order to generate outputs according to previously learned information patterns, following feedforward error-correction computations [75, 76]. Perhaps, such computational machinery makes the cerebellum's information processing more reliable, in accordance with the low variability in its DFC, although further research is needed to shed light on these findings.

**Declaration of Competing Interest** The authors declare that they have no competing interests.

# CRediT authorship contribution statement

**Izaro Fernandez-Iriondo:** Performed the analyses, Made the figures, Drafted the first manuscript, Wrote the manuscript. **Antonio Jimenez-Marin:** Performed the analysis, Made the figures, Wrote the manuscript. **Ibai Diez:** Preprocessed the images, Wrote the manuscript. **Paolo Bonifazi:** Wrote the manuscript. **Stephan P. Swinnen:** Acquired the data, Wrote the manuscript. **Miguel A. Muñoz:** Supervised the research, Equal last-author contribution, Wrote the manuscript. **Jesus M. Cortes:** Drafted the first manuscript, Supervised the research, Equal last-author contribution, Wrote the manuscript.

**Acknowledgements** A.J.M. acknowledges financial support from a predoctoral grant from the Basque Government (PRE_2019_1_0070). J.M.C. and P.B. acknowledge financial support from Ikerbasque (The Basque Foundation for Science) and from the Ministerio Economia, Industria y Competitividad (Spain) and FEDER (grant DPI2016-79874-R to J.M.C., grant SAF2015-69484-R to P.B.). J.M.C. acknowledges financial support from the Department of Economical Development and Infrastructure of the Basque Country (Elkartek Program, KK-2018/00032 and KK-2018/00090). S.P.S was supported by the the FWO Research Foundation Flanders (G089818N), the Excellence of Science funding competition (EOS; 30446199) and the KU Leuven Special Research Fund (grant C16/15/070). M.A.M acknowledges financial support from the Spanish Ministry and Agencia Estatal de investigación (AEI) through grant FIS2017-84256-P (European Regional Development Fund), as well as the Consejería de Conocimiento, Investigación y Universidad, Junta de Andalucía and European Regional Development Fund (ERDF), ref. SOMM17/6105/UGR for financial support.

# List of Tables and Figures

**Table 1**
**Brain anatomy for the three most relevant modules.** Module 1, formed by posterior cerebellar structures, had the highest SC-DFC similarity independent of the window lengths. Module 11 was one of the modules with higher DFC variability across different windows, formed by several cortical regions. Module 14, with a relative DFC variability across time windows, provided the lowest SC-DFC similarity independent of the window lengths and included the anterior cerebellar structures, brainstem and cortical regions. For these three modules, we only reported overlapping percentages bigger than 5%.

| Module 1 | Module 11 | Module 14 |
|---|---|---|
| Cerebelum_Crus1 (%21.879) | Calcarine (%19.2356) | BrainStem (% 15.718) |
| Cerebelum_8 (%19.1888) | Temporal_Mid (%15.601) | Cerebelum_6 (%15.2804) |
| Cerebelum_Crus2 (%15.4499) | Lingual (%14.3722) | Fusiform (%14.2614) |
| Cerebelum_9 (%7.8111) | Temporal_Inf (%10.07) | Cerebelum_4_5 (%14.031) |
| Cerebelum_6 (%7.0914) | Precuneus (%8.6727) | Lingual (%10.1075) |
|  | Cuneus (%6.1875) |  |

**Table 2**
**Functional and structural connectivity of modules 1 and 14 to the rest of the brain.** Note that the two modules connect both functionally and structurally to one another. Module 14 is also connected bilaterally to the anterior part of the paracentral lobule (2.65% of overlapping index), which is part of the supplementary motor cortex (see figure 5).

| FC (Module 1) | SC (Module 1) |
|---|---|
| Temporal_Inf (% 14.1851) | Temporal_Inf (% 12.9084) |
| Fusiform (% 12.2463) | Fusiform (% 11.7059) |
| Lingual (% 10.1058) | Lingual (% 9.4173) |
| Occipital_Mid (% 8.9656) | Cerebelum_6 (% 8.6846) |
| Cerebelum_6 (% 8.4921) | BrainStem (% 7.8873) |
| Cerebelum_4_5 (% 6.1129) | Cerebelum_4_5 (% 7.3657) |
| BrainStem (% 5.9853) | Temporal_Mid (% 6.8572) |
| Occipital_Inf (% 5.6888) | Occipital_Inf (% 5.5512) |

| FC (Module 14) | SC (Module 14) |
|---|---|
| Cerebelum_8 (% 7.0071) | Cerebelum_8 (% 7.5645) |
| Cerebelum_Crus1 (% 6.6914) | Cerebelum_Crus1 ( % 7.3713 ) |
| Temporal_Inf (% 5.8074) | Cerebelum_Crus2 (% 6.0049) |
| Calcarine (% 5.6633) | Temporal_Inf (% 5.3522) |
| Cerebelum_Crus2 (% 5.5094) | Temporal_Sup (% 5.08981) |
| Lingual (% 5.0259) |  |

**Table S1**
**Brain anatomy for modules 2-10, and 12-13.** As in Table 1, we only report overlapping percentages above 5%.

| Module 2 | Module 3 | Module 4 |
|---|---|---|
| Temporal_Sup (%7,869) | Parietal_Inf (%17,751) | Frontal_Mid (%28,969) |
| Insula (%6,666) | Postcentral (%16,609) | Precentral (%21,321) |
| Frontal_Inf_Orb (%5,346) | Precuneus (%14,837) | Postcentral (%15,076) |
| | Parietal_Sup (%12,303) | Frontal_Sup (%11,633) |
| | Precentral (%7,360) | SupraMarginal (%6,204) |
| | Occipital_Sup (%5,961) | Frontal_Inf_Tri (%5,553) |
| | Cuneus (%5,796) | |

| Module 5 | Module 6 | Module 7 |
|---|---|---|
| Supp_Motor_Area (%31,390) | Temporal_Sup (%80,328) | Frontal_Sup_Medial (%35,326) |
| Cingulum_Mid (%26,442) | Temporal_Mid (%17,213) | Frontal_Mid (%22,417) |
| Precentral (%12,536) | | Frontal_Sup (%20,520) |
| Frontal_Mid (%8,967) | | Cingulum_Ant (%13,832) |
| Frontal_Sup (%7,7866) | | |
| Paracentral_Lobule (%5,678) | | |

| Module 8 | Module 9 | Module 10 |
|---|---|---|
| Precuneus (%19,252) | Frontal_Inf_Tri (%10,092) | Temporal_Inf (%14,276) |
| Angular (%11,769) | Rectus (%8,978) | Temporal_Mid (%9,564) |
| Parietal_Sup (%10,814) | Frontal_Med_Orb (%8,673) | Temporal_Pole_Sup (%7,899) |
| Occipital_Mid (%9,191) | Frontal_Sup_Orb (%7,864) | Insula (%7,663) |
| Parietal_Inf (%9,073) | Frontal_Inf_Orb (%7,612) | Hippocampus (%6,684) |
| Occipital_Sup (%7,324) | Cingulum_Ant (%5,630) | ParaHippocampal (%6,195) |
| SupraMarginal (%6,579) | Frontal_Mid (%5,590) | Fusiform (%6,061) |
| Postcentral (%6,002) | | Temporal_Sup (%5,082) |
| Cuneus (%5,642) | | |

| Module 12 | Module 13 |
|---|---|
| Temporal_Mid (% 27,226) | Temporal_Inf (%73,469) |
| Occipital_Mid (%14,608) | |
| Temporal_Sup (%13,273) | |
| SupraMarginal (% 6,720) | |
| Postcentral (%5,376) | |

**Table S2**
**SC-DFC similarity for** $\delta = \mathcal{T}$, considering only one time-window with a length equal to the entire time-series. In this case, the DFC variability could not be calculated, as it was defined by averaging different window pairs. Note that for $\delta = \mathcal{T}$, modules 1 and 14 also had extreme values with respect to SC-DFC similarity.

| Module | SC-DFC similarity, $r^m$ |
| --- | --- |
| 14 | 0.42 |
| 10 | 0.53 |
| 5 | 0.53 |
| 9 | 0.55 |
| 2 | 0.56 |
| 4 | 0.60 |
| 7 | 0.60 |
| 3 | 0.61 |
| 8 | 0.62 |
| 11 | 0.66 |
| 12 | 0.67 |
| 1 | 0.77 |

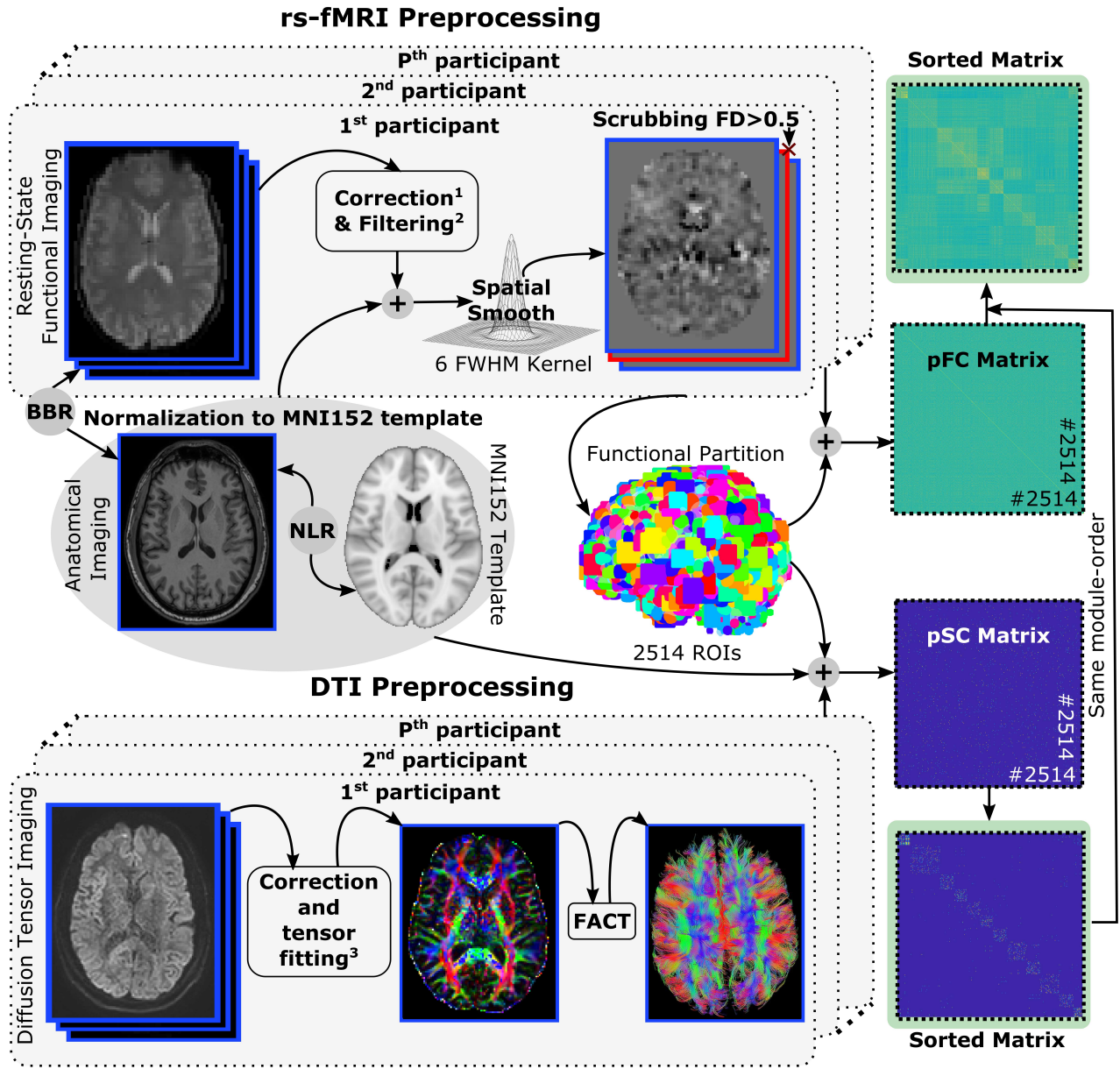

[1] rs-fMRI correction: slice-time correction, head-movement correction and intensity normalization
[2] rs-fMRI filtering: regress out of cofounding factors (movement, FD, CSF, WM and w/o GSR) and bandpass filter
[3] DTI correction and tensor fitting: eddy current and motion correction; and diffusion tensor fitting

**Figure 1: Neuroimaging preprocessing pipeline.** Three acquisitions were obtained for each subject: High-resolution anatomical images (T1), functional images at rest (fMRI) and diffusion tensor imaging (DTI). Following a state-of-the-art neuroimaging preprocessing pipeline (summarized in footnotes 1-3), we obtained time series of the blood oxygenation level-dependent (BOLD) signal for each network node, defined by a functional partition of 2,514 regions covering the entire brain and brainstem. Using the same network nodes, we also built structural connectivity matrices by counting the number of streamlines between pairs of nodes, thereby obtaining one connectivity class per subject that we averaged to achieve population matrices (pSC, pFC). Finally, for the comparison of pSC and pFC, we re-ordered the latter according to the results after modularizing pSC. Here, pFC refers to any generic time-window. Abbreviations: FD = Frame Displacement; CSF = Cerebrospinal Fluid; WM = White Matter; GSR = Global Signal Regression; BBR = Boundary-Based Registration; NLR = Non-Linear Registration.

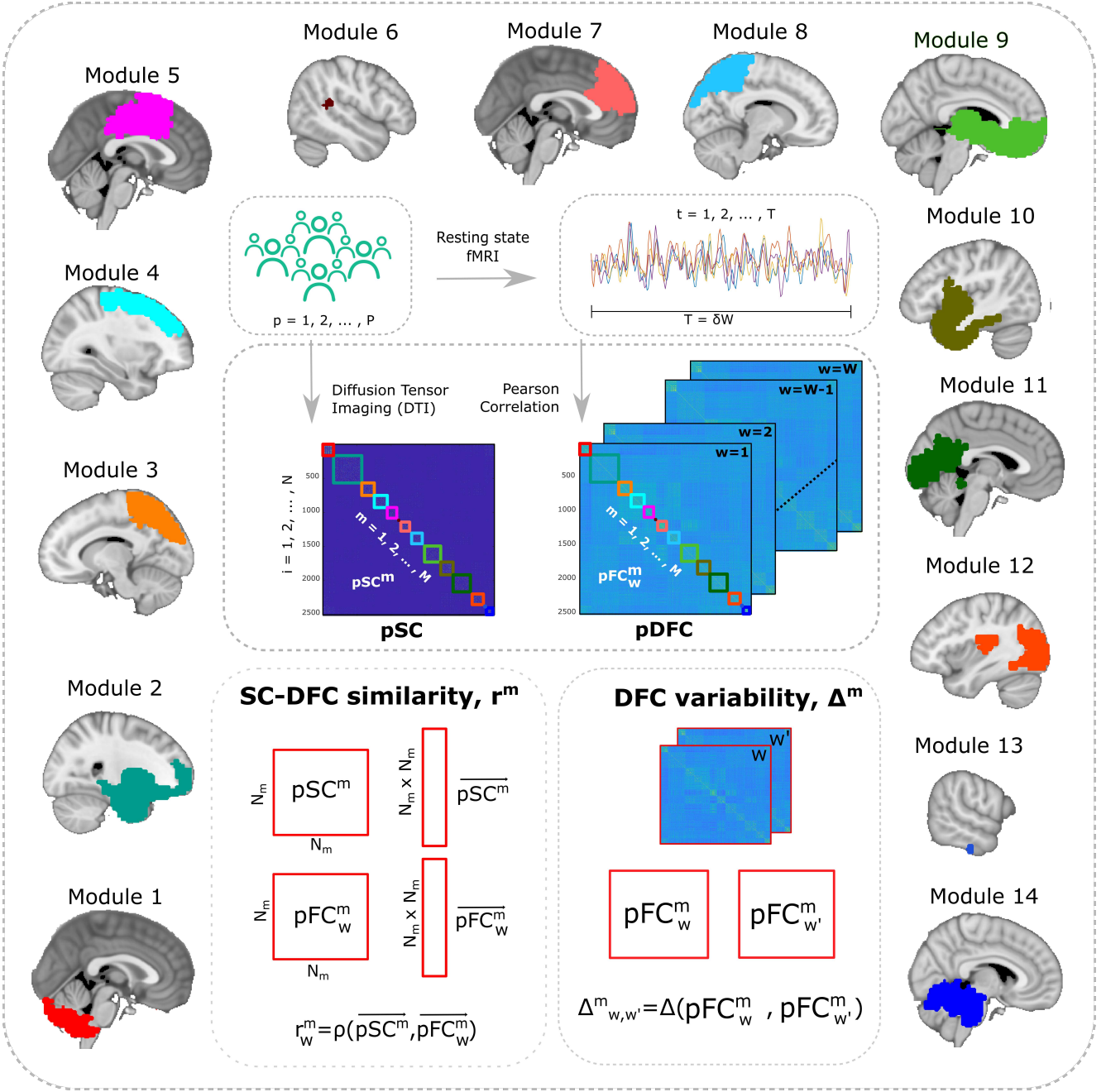

Figure 2: Scheme to assess SC-DFC similarity and DFC variability at the level of the modules. For each participant $p$, we obtained one SC matrix and a sequence of $W$ matrices of $FC_w$, that were averaged to obtain the population pSC and pFC matrices. After maximization of modularity in pSC, we obtained M=14 modules, marked by different colored squares and represented in the brain with one representative slide. All $pFC_w$ matrices were reordered using the structural modules and then, the SC-DFC similarity was estimated within each module $m$ separately, by calculating the Pearson correlation between vector-wise representations of the $pSC^m$ and $pFC_w^m$ matrices, and finally averaging over the windows $w$. The DFC variability along different time windows was assessed by calculating the pairwise spectral distance between matrices $FC_w^m$ and $FC_{w'}^m$, and finally averaging over window pairs $w$ and $w'$. Modules 1, 5, 7, 9, 11 and 14 were bilateral; Modules 2, 4, 6, 8 and 13 were mainly located in the right hemisphere; Modules 3, 10 and 12 were mainly located in the left hemisphere. The sizes of the modules $N_m$, measured in number of network nodes per module, were 176, 404, 197, 168, 180, 2, 171, 191, 251, 191, 262, 189, 1 and 131, respectively.

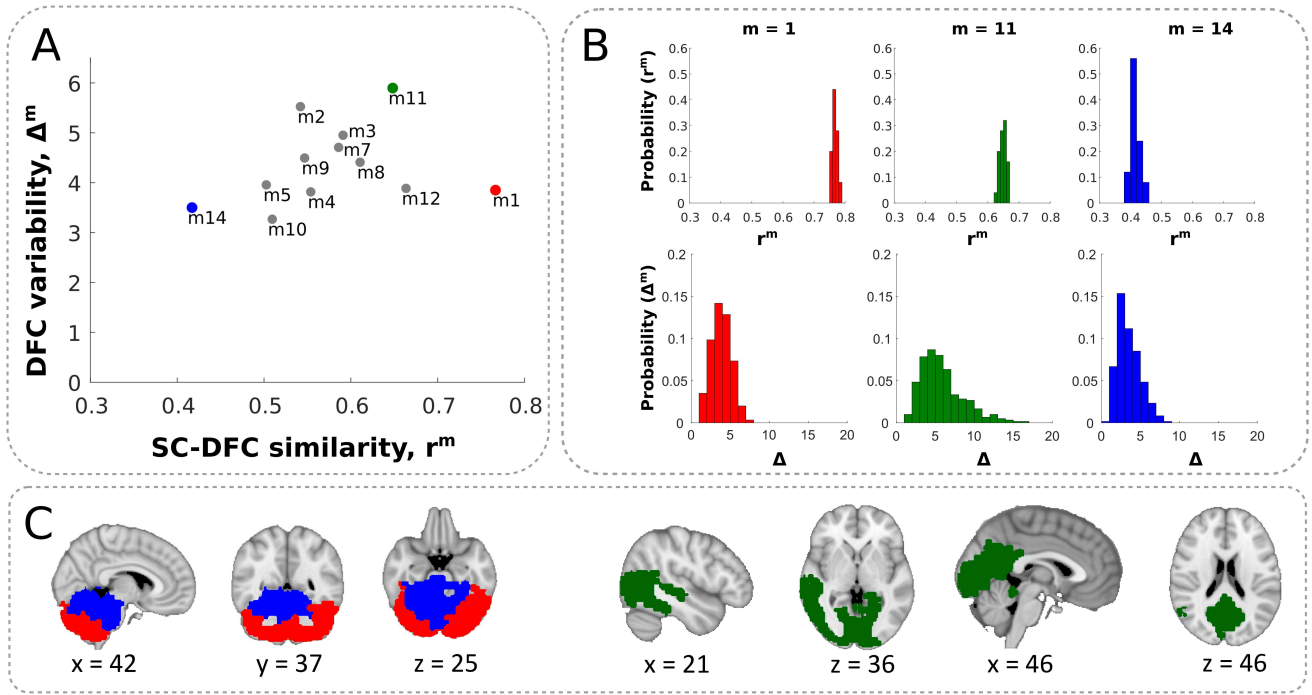

Figure 3: **The amount of DFC variability and SC-DFC similarity define three different module-scenarios.** A: For all modules, we plotted their representation in the plane ($r^m, \Delta^m$), from which we detected three different scenarios: Module 1 (with the highest $r^m$ value, red), module 11 (with the highest $\Delta^m$ value, green) and module 14 (with the lowest $r^m$, blue). The points represent the average values of $r^m$ over the time windows. Similarly, $\Delta^m$ is the average of the window pairs. B: Probability distribution of all the $r^m$ and $\Delta^m$ values obtained for the different windows. C: Anatomical representation of three modules, where x, y, z represents the slice number in each axis. A,B,C: Results for a window length of $\delta=8$, which for non-overlapping windows and a total number of 200 time points resulted in 25 different windows over which the matrices $FC_w^m$ were calculated.

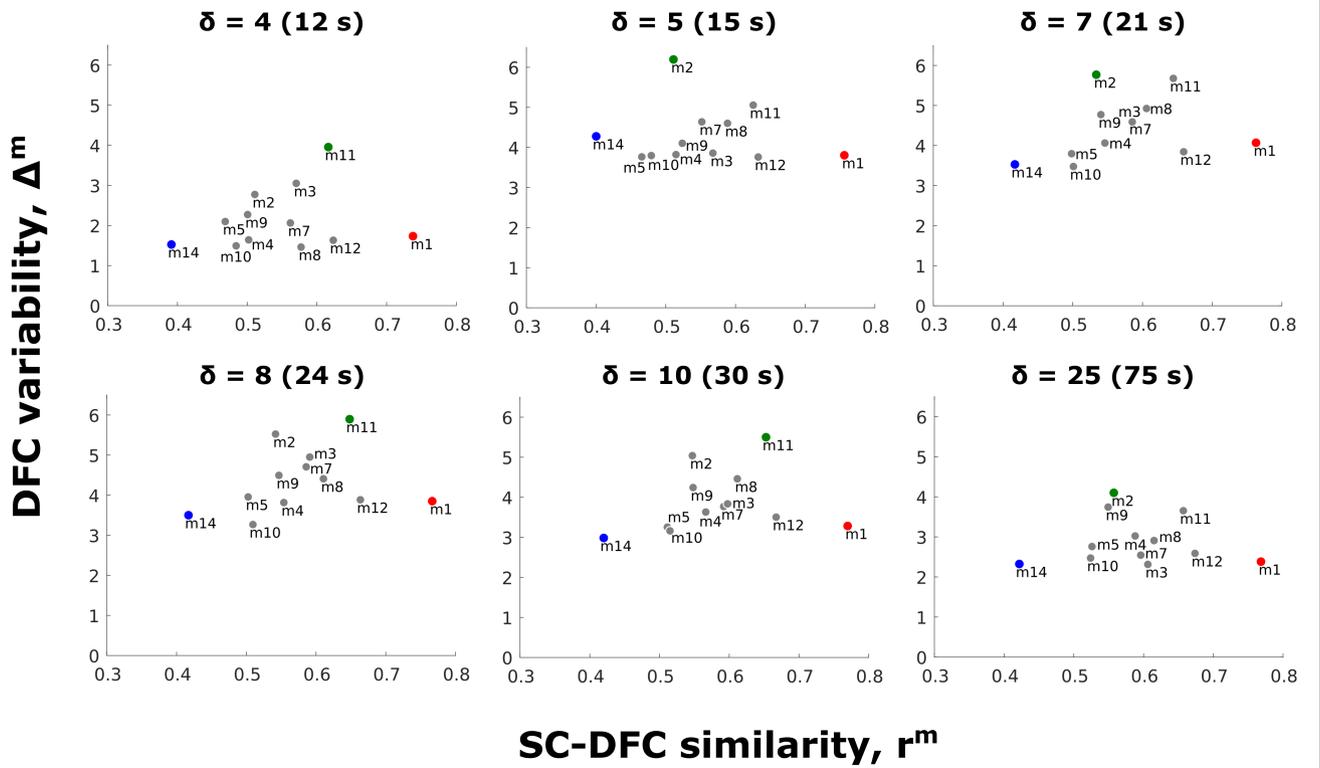

**Figure 4: Robustness of the three relevant module-scenarios for different window lengths.** The characteristics of modules 1 and 14 have the highest and lowest value of $r^m$, respectively, whilst keeping low $\Delta^m$, preserved independently on the value of window length $\delta$. However, module 11 in figure 3 had the highest $\Delta^m$ value and when changing $\delta$, the roles switch between module 2 and 11. Importantly, the two invariant modules 1 and 14 are both parts of the cerebellum. The window length $\delta$ is given in time points, each one corresponding to a time duration of TR = 3 sec.

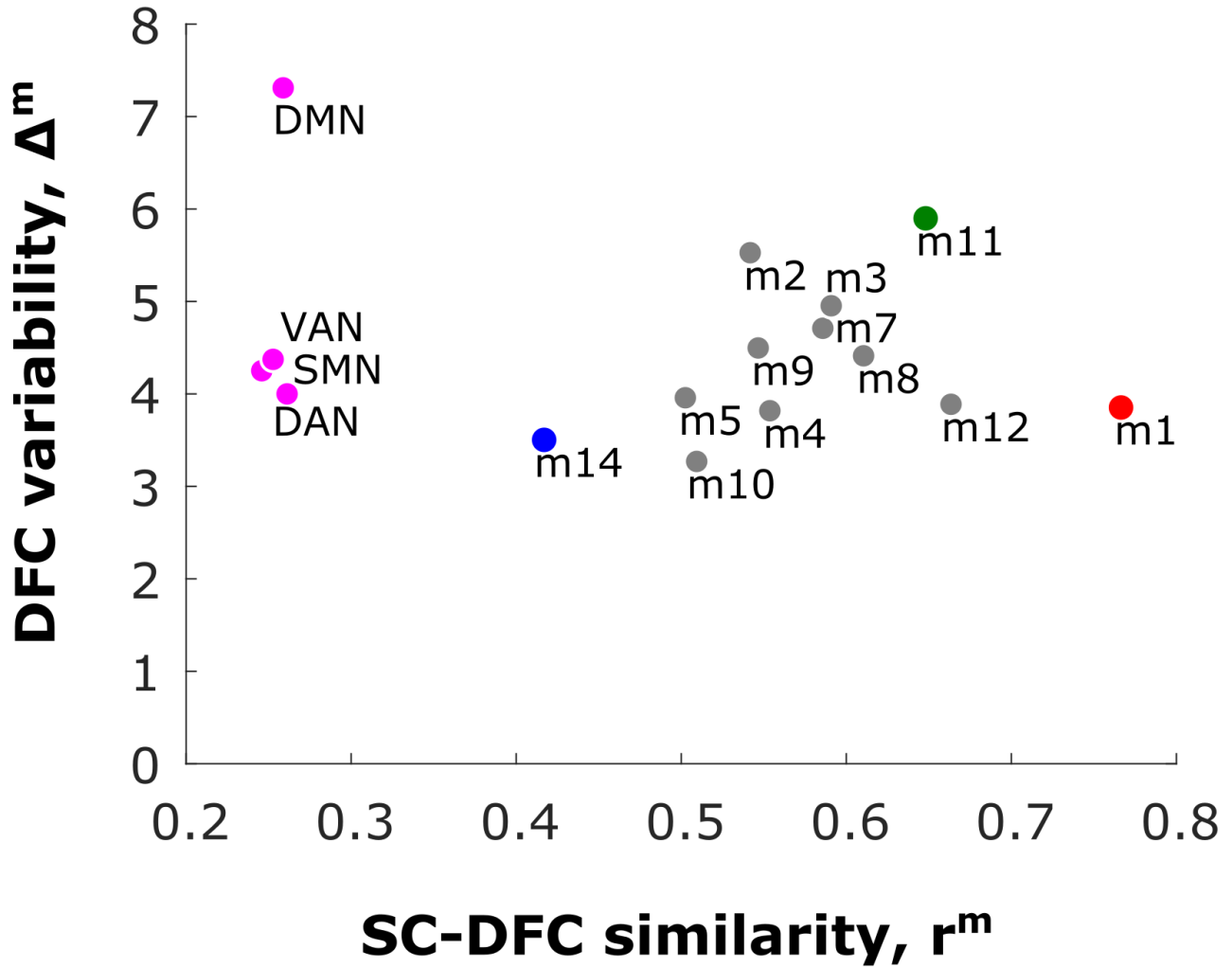

**Figure 5: Structure-function modules vs classical resting state networks (RSNs).** Modules m1-m14 were obtained from the structural connectivity matrix. By contrast, RSNs are purely functional modules, as reflected by the weak SC-DFC similarity (pink). Moreover, the DMN had the largest DFC variability, which suggests different computational principles for this network (not really constrained to brain structure). Abbreviations: Default Mode Network (DMN), Ventral Attention Network (VAN), Somatomotor Network (SMN), Dorsal Attention Network (DAN).

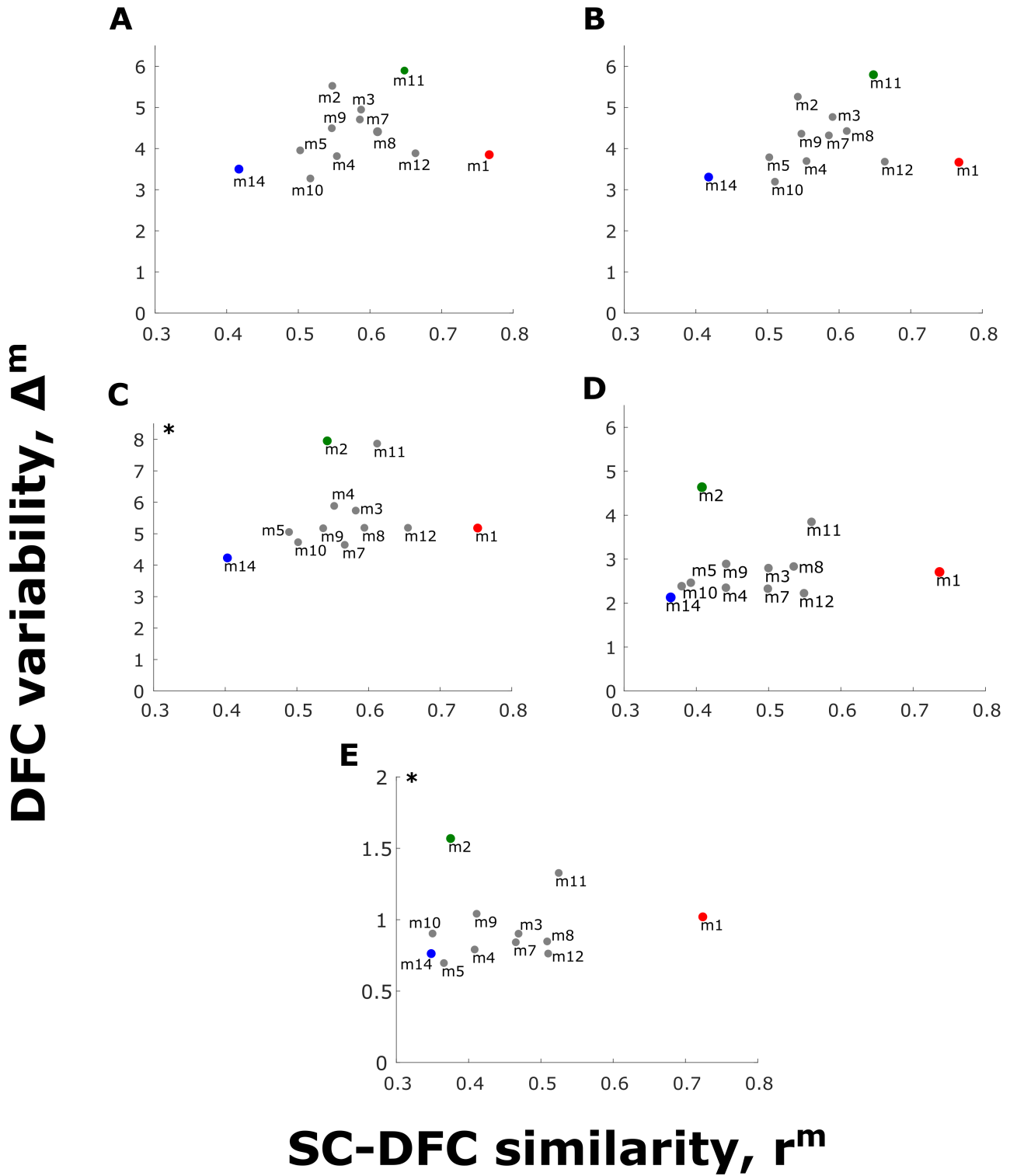

**Figure S1: The roles of modules 1 and 14 are preserved in different control conditions,** when calculating SC matrices by probabilistic tractography (**A**), with 96%-overlapping sliding windows (**B**), without GSR removal (**C**), and calculating FC matrices using the instantaneous amplitude of the complex analytical signal (**D**), and the instantaneous phase, also known as the phase locking value (**E**). (**A-E**): In all panels, module 1 had the highest SC-DFC similarity and low DFC variability values, while module 14 had the lowest SC-DFC similarity value and also low DFC variability values. Strikingly, module 1 and 14 were localized in the posterior and anterior parts of the cerebellum, respectively. (**C,E**): * different scale.

Small variation in dynamic functional connectivity in cerebellar networks

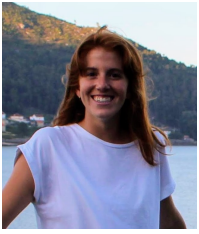

Izaro Fernandez-Iriondo received her Bachelor's degree in Physics in 2019 from the University of the Basque Country (UPV). She is currently studying a master's degree in computer engineering and intelligent systems at the UPV in San Sebastian, aiming to complete this with a machine learning project applied to structural and functional brain networks.

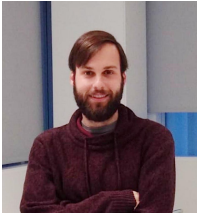

Antonio Jimenez-Marin is a PhD student at the Biomedical Research Program of the University of the Basque Country. His research is carried out in the Computational Neuroimaging Lab at Biocruces Bizkaia Health Research Institute. He obtained his degree in Telecommunication Technology Engineering at the University of Granada (2015) and his MSc in Biomedical Engineering at the University of the Basque Country (2018). His research interests focus on brain connectivity in health and disease.

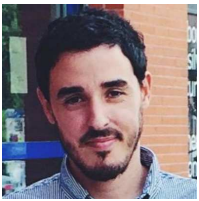

Ibai Diez received his Bachelor's degree in Telecommunication Engineering in 2009 from Deusto University, Spain, and his PhD in 2015 from University of the Basque Country (UPV/EHU, Spain). He is currently a postdoctoral researcher in the Neurology Department at Massachusetts General Hospital – Harvard Medical School. His research interest include structural and functional brain connectivity, biomedical data analysis and functional integration in the brain.

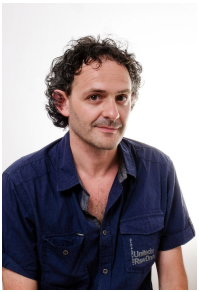

Paolo Bonifazi is a Tenured Ikerbasque Researcher at the Biocruces-Bizkaia Health Research Institute in Bilbao (Spain), within the Computational Neuroimaging Lab. He is a System Neuroscientist with a degree in Physics (university of Perugia, Italy) and a PhD in Neuroscience (SISSA, Trieste, Italy). After completing two postdoctoral positions in the UK (with Prof. Hugh Robinson) and France (with Dr Rosa Cossart), he became a Research Associate at the Tel Aviv University in the group of Prof. Ari Barzilai and the late Prof. Eshel Ben-Jacob. His research interests are multi-disciplinary, focusing on the structure and function of brain circuits, spanning from microcircuits to brain networks, bridging experimental and computational approaches, the latter inspired by complex networks.

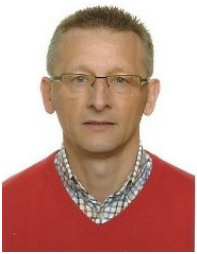

Stephan P. Swinnen started his research career in movement control at the University of California at Los Angeles (UCLA, 1983-85) under the direction of Prof. R. A. Schmidt. He completed his PhD at KU Leuven in 1987 and he was awarded a Francqui Research Professor position (2013-2016). He currently directs the Movement Control & Neuroplasticity Research Group at KU Leuven (Belgium). His research interests focus on mechanisms underlying movement control and neuroplasticity in normal and pathological conditions, using multidisciplinary approaches focusing on the study of brain function, structure, connectivity and neurochemicals.

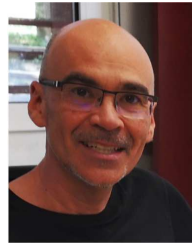

Miguel A. Muñoz is a Full Professor in Physics at the University of Granada (Spain). He is an expert in statistical mechanics and has worked on, among other issues, non-equilibrium phase transitions, critical and collective phenomena and stochastic processes. In particular, he helped develop the theory of "self-organized criticality" and approaches towards the study of non-equilibrium phenomena, network theory and complex systems. His research interests span from fundamental principles of statistical mechanics to interdisciplinary problems in evolutionary biology, theoretical ecology and systems neuroscience.

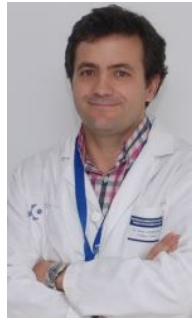

Jesus M. Cortes is an Ikerbasque Research Professor and the head of the Computational Neuroimaging Lab at the Biocruces-Bizkaia Health Research Institute in Bilbao (Spain). He teaches Brain Connectivity and Neuroimaging on the M.Sc. of Biomedical Engineering. He obtained a Ph.D. in Physics in 2005 and completed three postdoctoral positions in The Netherlands (supervised by Prof. Bert Kappen), USA (supervised by Prof. Terry Sejnowski) and UK (supervised by Prof. Mark van Rossum). Among many other merits, he participated in the team headed by Prof. Mazahir T. Hasan that achieved the milestone of receiving the first project funded in Spain by the Brain Initiative. He also got Ph.D. with Distinction. His area of research now focuses on brain connectivity, neuroimaging and machine learning methods applied to healthy and pathological conditions.